\documentclass[english,prd,twocolumn,superscriptaddress,nofootinbib,preprintnumbers]{revtex4-1}
\usepackage[latin1]{inputenc}
\usepackage{graphicx}
\usepackage{amssymb}
\usepackage{color}
\usepackage{float}
\usepackage{amsmath}
\usepackage{amsfonts}
\usepackage{dcolumn}
\usepackage{hyperref}
\usepackage{subfigure}
\usepackage{soul,xcolor}

\newcommand{\be}{\begin{equation}}
\newcommand{\ee}{\end{equation}}
\newcommand{\bea}{\begin{eqnarray}}
\newcommand{\eea}{\end{eqnarray}}
\newcommand{\lb}{\label}

\begin{document}

\title{On the consistency of the expansion with the perturbations}

\author{Radouane Gannouji}
\affiliation{Instituto de F\'{\i}sica, Pontificia Universidad  Cat\'olica de Valpara\'{\i}so, 
Casilla 4950, Valpara\'{\i}so, Chile}
\author{David Polarski}
\affiliation{Laboratoire Charles Coulomb, Universit\'e Montpellier 2 \& CNRS UMR 5221, 
F-34095 Montpellier, France}

\begin{abstract}
Assuming a simple form for the growth index $\gamma(z)$ depending on two parameters 
$\gamma_0\equiv \gamma(z=0)$ and $\gamma_1\equiv \gamma'(z=0)$, we show that these parameters 
can be constrained using \emph{background} expansion data. We explore systematically the 
preferred region in this parameter space. Inside General Relativity we obtain that models 
with a quasi-static growth index and $\gamma_1\approx -0.02$ are favoured. We find further
the lower bounds $\gamma_0\gtrsim 0.53$ and $\gamma_1\gtrsim -0.15$ for models inside GR. 
Models outside GR having the same background expansion as $\Lambda$CDM and 
arbitrary $\gamma(z)$ with $\gamma_0 = \gamma_0^{\Lambda CDM}$, satisfy $G_{{\rm eff},0}>G$ 
for $\gamma_1 > \gamma_1^{\Lambda CDM}$, and $G_{{\rm eff},0}<G$ for 
$\gamma_1 < \gamma_1^{\Lambda CDM}$. The first models will cross downwards 
the value $G_{{\rm eff}}=G$ on very low redshifts $z<0.3$, while the second models 
will cross upwards $G_{{\rm eff}}=G$ in the same redshift range. This makes the realization of 
such modified gravity models even more problematic.  
\end{abstract}

\maketitle

\section{Introduction}
Understanding the origin of the present accelerated expansion of the universe remains 
a challenge 
for theorists. 
A huge number of theoretical models and mechanisms were suggested and investigated that can 
produce this late-time accelerated expansion, see the reviews \cite{SS00}. 
It is remarkable that the simplest model where gravity is described by General Relativity 
(GR) containing a cosmological constant $\Lambda$ offers broad consistency with existing data, 
especially on large cosmic scales. 
While deriving the tiny value of $\Lambda$ from first principles using quantum field theory 
still remains an outstanding problem, the phenomenological agreement of this model with 
observations provides a benchmark for the assessment of other proposed dark energy (DE) 
models. 

An efficient way to make progress is to carefully explore the phenomenology 
of the proposed models and to compare it with observations \cite{WMEHRR13}. 
Hence it is important to find tools which can efficiently discriminate between models, 
or between classes of models (e.g. \cite{SSS14}). 
The growth index $\gamma$, which gives a way to parametrize the 
growth of density perturbations of the non-relativistic matter (dust) component is an 
interesting example of such a phenomenological tool.
This approach was pioneered long ago in order to discriminate 
spatially open from spatially flat universes \cite{P84} and then generalized to other 
cases \cite{LLPR91}. 
It was revived recently in the context of dark energy models \cite{LC07} and it has 
been investigated and used in various disguise (see e.g. \cite{gamma}). 
As it is the case with many other quantities of interest, we can expect a significant 
improvement of the measurement of $\gamma$ in the future thereby providing new observational 
constraints on DE models.
A crucial property is that the growth index has a clear signature when DE reduces to a 
cosmological constant $\Lambda$: the growth index at very low redshifts lies around $0.55$ 
and it is quasi-constant. This behaviour can be extended to noninteracting DE models inside 
GR with a constant (or even smoothly varying) equation of state $w_{DE}$, while a strictly 
constant $\gamma$ is very peculiar \cite{PG07}.
Such behaviour is strongly violated in some models beyond GR, see e.g. \cite{GMP08,MSY10} 
offering therefore the additional possibility to single out DE models formulated outside GR.

To constrain DE models, one can use the consistency of the background expansion with the 
matter perturbations growth. The growth index is just one of the phenomenological tools for 
the study of matter perturbations.
There are several ways in which it can be used in order to constrain DE models. 
One can assume some DE model and study the behaviour of the growth index together with the 
possible background expansions. Then the behaviour of $\gamma$ which is found expresses 
automatically the consistency mentioned above.  
Another way to exploit this consistency is by reconstructing the background 
expansion using the perturbations for a given class of DE models, a property emphasized 
some time ago \cite{S98}.
In principle, even inside GR, it requires the knowledge of the perturbation functions 
$\delta_m(z)$, and of some additional cosmological parameters, in order to reconstruct $H(z)$. 
So an exact reconstruction is generally a complicated problem. 

As we will show however, the growth index provides a very effective tool in this respect too. 
Actually, the reconstruction of $h(z)\equiv \frac{H(z)}{H_0}$ was given in \cite{PSG16} for 
a constant $\gamma$ inside GR. Here, we will extend this result to more general behaviours 
of $\gamma(z)$. We will further  extend this approach to modified gravity DE models and 
reconstruct the (effective) gravitational constant.
It is this use of the growth index that we address in the present work.   

\section{The growth index}
We recall briefly the basic equations and concepts concerning the growth index.
We consider a spatially flat Friedmann-Lema\^itre-Robertson-Walker (FLRW) universe 
filled with standard dust-like matter and DE components. We can neglect radiation in the 
matter and DE dominated stages.
Deep inside the Hubble radius, the evolution of linear scalar (density) perturbations 
$\delta_m =\delta\rho_m/\rho_m$ in the (dust-like) matter component follows from the 
equation (for GR)
\be
{\ddot \delta_m} + 2H {\dot \delta_m} - 4\pi G\rho_m \delta_m = 0~,\label{del}
\ee
where $H(t)\equiv \dot a(t)/a(t)$ is the Hubble parameter and $a(t)$ is the scale factor, 
while $G$ is Newton's gravitational constant. The evolution of the Hubble parameter as a function of the redshift $z=\frac{a_0}{a}-1$ at 
$z\ll z_{eq}$ reads 
\begin{align}
h^2(z) = \Omega_{m,0} (1+z)^3 + (1 - \Omega_{m,0}) e^{3\int_{0}^z \frac{1+w_{DE}(z')}{1+z'}dz'}\lb{h2z}
\end{align}
with $h(z)\equiv \frac{H}{H_0}$ and $w_{DE}(z)\equiv p_{DE}(z)/\rho_{DE}(z)$. Equation 
\eqref{h2z} holds for all non-interacting DE models inside GR. We have the useful relation
$w_{DE} = \frac{1}{3(1-\Omega_m)}~\frac{d\ln \Omega_m}{d\ln a}$ 
using the standard definition 
$\Omega_m = \Omega_{m,0} \left(\frac{a_0}{a}\right)^3 h^{-2}$.

Instead of $\delta_m$, it may be convenient to introduce the 
growth function  $f\equiv \frac{d \ln \delta_m}{d \ln a}$. Then \eqref{del} leads to the 
following nonlinear first order equation \cite{WS98}
\be
\frac{df}{dN} + f^2 + \frac{1}{2} \left(1 - \frac{d \ln \Omega_m}{dN} \right) f = 
                              \frac{3}{2}~\Omega_m~,\lb{df}
\ee
with $N\equiv \ln a$. The quantity $\delta_m$ is easily recovered from $f$, viz.  
\be
\delta_m(a) = \delta_{m,i}~\exp \left[ \int_{a_i}^{a} f(x') \frac{dx'}{x'} \right]~.
\ee
Obviously $f=p$ for $\delta_m\propto a^p$ (with $p$ constant). In particular $f\to 1$ in 
$\Lambda$CDM for large $z$ and $f=1$ in the Einstein-de Sitter universe. 
In order to characterize the growth of perturbations, the parametrization 
$f = \Omega_m(z)^{\gamma}$ has been intensively used and investigated 
in the context of dark energy, where $\gamma$ is the growth index. In general 
however, $\gamma$ is not constant and one should write 
\be
f=\Omega_m(z)^{\gamma(z)}~. \lb{gammagen} 
\ee 
Surprisingly, it turns out that the growth index is quasi-constant for $\Lambda$CDM. 
Such a behaviour holds also for smooth non-interacting DE models inside GR when 
$w_{DE}$ is constant \cite{PG07}. 
It is known however that this behaviour changes substantially in modified 
gravity, an important motivation for the use of the growth index in the study of DE. 
In many DE models outside GR the dynamics of matter perturbations is modified by 
the replacement $G\to  G_{\rm eff}$ in \eqref{del}, see e.g. \cite{BEPS00}, 
where $G_{\rm eff}$ is a model-dependent effective gravitational coupling.
Introducing the quantity
\be
g\equiv \frac{G_{\rm eff}}{G}\lb{g}~,
\ee  
we obtain instead of Eq. \eqref{df}
\be
\frac{df}{dN} + f^2 + \frac{1}{2} \left(1 - \frac{d \ln \Omega_m}{dN} \right) f = 
                              \frac{3}{2}~g~\Omega_m~.\lb{dfmod}
\ee
which can be recast into
\be
2 \ln \Omega_m~\frac{d\gamma}{dN} + (2\gamma-1)~\frac{d\ln \Omega_m}{dN} 
                              + 1 + 2\Omega_m^{\gamma} - 3 g \Omega_m^{1-\gamma} = 0~. 
\label{eq:all3}
\ee 
Note that in \eqref{dfmod}, \eqref{eq:all3}, the cosmological parameters 
$\Omega_i = \frac{8\pi G \rho_i}{3 H^2}$ are defined as in GR i.e. using Newtons 
gravitational constant $G$.
We also see a degeneracy which can be read from equation \eqref{eq:all3}. In fact, 
we can have an infinite number of combinations $(g,\gamma)$ which produce the same 
$\Omega_m$ as for example in $\Lambda$CDM. It is interesting that $G_{\rm eff}$ can be constructed in an algebraic way once the background and the linear perturbations are measured with enough precision. Note that $G_{\rm eff}$ can also be scale dependent in modified gravity models with screening 
of a fifth force on small scales. This in turn induces a scale dependence of $\gamma$. 
In that case all equations and results hold for each scale separately. In this work we will 
consider a growth index which is essentially scale independent.  
From \eqref{dfmod} one gets the following equality 
\begin{align}
w_{DE} &= - \frac{1}{3(2\gamma -1)} ~\frac{2 \frac{d\gamma}{dN}\ln \Omega_m  + 
                       1 + 2\Omega_m^{\gamma} - 3 g \Omega_m^{1-\gamma}}{1-\Omega_m} \lb{wgen}\\
   & \equiv - \frac{1}{3(2\gamma -1)} ~
                        \left[ \frac{2 \frac{d\gamma}{dN}~\ln \Omega_m }{1-\Omega_m} 
                       + F(\Omega_m,~\gamma,~g) \right]~.  \lb{F}
\end{align}
which expresses the essential physical content of our formalism. The last expression defines 
the quantity $F(\Omega_m,~\gamma,~g)$ which encodes the dependence of $w_{DE}$ on $\Omega_m$ 
for constant $\gamma$. The case $g=1$ reduces to GR. We refer to \cite{PSG16} for additional 
details. 

\section{Reconstruction}
The basic formalism outlined in the previous section allows for a reconstruction program in 
various ways. If we include modified gravity DE models, we have three unknown functions 
$h(z)$, $\gamma(z)$ and $g(z)$ in (\ref{eq:all3}). Fixing two of these functions, or making 
reasonable assumptions, one can reconstruct the third unknown function. 
\subsection{Reconstruction of the background expansion inside GR}
Let us consider first DE models where gravity is described by GR $(g=1)$. As noted some 
time ago, it 
is interesting that the background expansion can be reconstructed from the matter 
perturbations \cite{S98}. Hence knowing both the expansion and the perturbations growth one 
can check the consistency of a given model. 
As it was emphasized in \cite{PSG16}, for 
non-interacting DE models, when the growth index $\gamma$ is constant, this 
mathematical property reduces to the fact that all background quantities can be expressed in 
parametric form using the variable $\Omega_m$. One obtains in particular \cite{PSG16} 
from (\ref{eq:all3}) for constant $\gamma$
\be
\ln (1+z) = (2\gamma-1) \int_{\Omega_{m,0}}^{\Omega_m} \frac{d\ln\Omega_m}
                     {1+2\Omega_m^{\gamma}-3\Omega_m^{1-\gamma}}~. \lb{zOm}
\ee 
The cosmic time $t$ can be expressed in a similar way \cite{PSG16}.   
From \eqref{zOm}, one can recover $h(z)$ and reconstruct therefore the background expansion.

This result can be extended to a larger number of models for which the growth index is 
not exactly constant. Indeed, even in this case it is still possible to find $\Omega_m(z)$ 
by solving the eq. (\ref{eq:all3}). Clearly, this equation is useful only provided we know $\gamma(z)$ or at least if we 
can make simple assumptions concerning its behaviour. 
Ideally, it would be very useful if we can describe the functional dependence of $\gamma(z)$ 
with a limited set of parameters.  
We can write in full generality a Taylor expansion around its value today, viz.
\begin{align}
\gamma &= \gamma_0 +\gamma_1~(1-x) + \gamma_2~(1-x)^2 + ...\\
&=\gamma_0 + \gamma_1~\frac{z}{1+z} + \gamma_2~\Bigl(\frac{z}{1+z}\Bigr)^2 + ...\lb{Tx}
\end{align}
with $x\equiv \frac{a}{a_0}$.
Obviously, it is desirable to have only two parameters when we derive observational constraints.
Hence, instead of \eqref{Tx} we will use the more tractable representation
\be
\gamma = \gamma_0 + \gamma_1~(1-x)~. \lb{Tx1}
\ee
On one hand, this choice is motivated by the fact that $\gamma$ is quasi-constant for a 
large class of models inside GR, and for $\Lambda$CDM in the first place.  
For these models it is clear that accurate fits are obtained already with \eqref{Tx1}. 
Hence for these models, \eqref{Tx1} provides a fit linear in $a$ valid 
in the full range probed by the observations, and actually everywhere. This is in the same 
spirit as the CPL parametrization of the equation of state (EoS) parameter $w_{DE}$ \cite{CP01}.
On the other hand, \eqref{Tx1} holds for any model provided it is used on small enough 
redshifts.   

Let us return to non interacting DE models with constant, or smoothly varying, $w_{DE}$. 
In that case, the behaviour of $\gamma$ up to redshifts of a few is very well approximated 
with \eqref{Tx1} (see e.g. \cite{PSG16}). 
For our purpose, we can make it more quantitative and we will say that a fit is good provided 
the reconstructed expansion is accurate. 
In other words, when the fit \eqref{Tx1} is substituted in \eqref{eq:all3}, with 
the \emph{true} parameters $\gamma_0$ and $\gamma_1$, the recontructed $h(z)$ should 
very close to the \emph{true} function $h(z)$. 

\begin{figure}
\begin{centering}
\includegraphics[scale=.8]{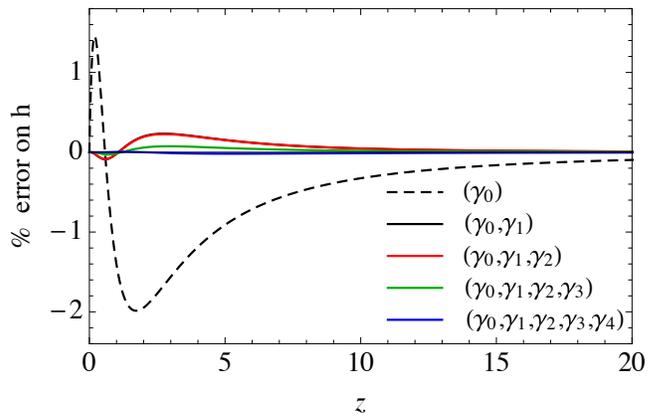}
\par\end{centering}
\caption{The reconstructed relative Hubble function $h(z)$ is shown for our fiducial 
$\Lambda$CDM model with $\Omega_{m,0}=0.30$ using 
$\gamma_0^{\Lambda CDM},~\gamma_1^{\Lambda CDM}$ in the expansion \eqref{Tx1}. 
Even the approximation $\gamma =\gamma_0^{\Lambda CDM}$ induces no more than a 
$2\%$ error. With the first order expansion \eqref{Tx1}, the error reduces to about $0.2\%$. The expansion to first order (black) is hardly distinguishable from the second order expansion (red). Inclusion of higher order terms yields a marginal improvement in the accuracy. }  
\label{fig1}
\end{figure}

We illustrate our results with the fiducial $\Lambda$CDM model, see figure \ref{fig1}.
The reconstruction of $h(z)$ turns out to be remarkably accurate already 
when the first order expansion \eqref{Tx1} is used, with errors less than $0.2\%$. 
As we can see further from figure \ref{fig1}, it is interesting that inclusion of 
the next orders in the expansion barely improves the accuracy. 
Of course, this accuracy is not related to observational 
uncertainties.

To summarize, the expansion \eqref{Tx1} up to first order provides a remarkably accurate 
reconstruction of the background expansion rate. 

Actually, even the zeroth order, that is if we approximate $\gamma$ by its present value 
$\gamma_0$, gives a good reconstruction with a maximal error of about $2\%$ only on the 
redshift range $0\lesssim z\lesssim 3$. However an inaccurate reconstruction can easily 
lead to false conclusions. We see from figure \ref{fig1} that DE appears to be partly of 
the phantom type showing a ``phantom-divide'' crossing on some low redshift. Taken at face 
value it could lead to the conclusion that quintessence models are ruled out. 
Of course, this is because we have taken $\gamma_0$ and $\gamma_1$ corresponding 
to the peculiar case of $\Lambda$CDM so the slightest inaccuracy can lead to a phantom 
behaviour. When the first order expansion \eqref{Tx1} is used, this phantom-divide crossing 
disappears essentially. 

\begin{figure}
\includegraphics[scale=.7]{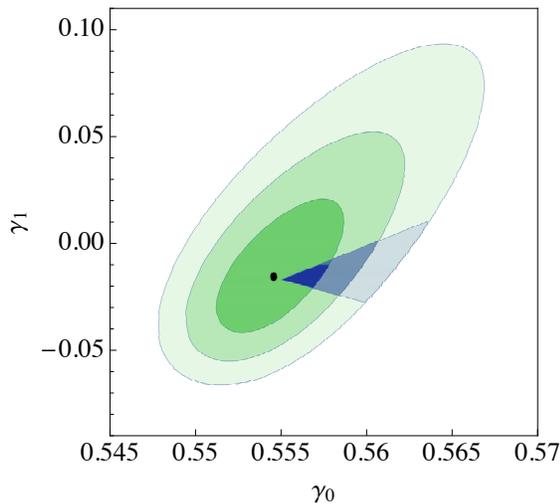}
\caption{The favoured region in the $(\gamma_0, \gamma_1)$ plane is shown when GR is 
assumed $(g=1)$ and $\Omega_{m,0}=0.3$ using SNIa data. The quantity $\gamma_0$ is rather sharply constrained 
at $2\sigma$, $0.549\lesssim \gamma_0 \lesssim 0.562$. In contrast, the constraint on 
$\gamma_1$ is much looser, $-0.06\lesssim \gamma_1\lesssim 0.05$.     
The best fit is $\gamma_0=0.555$ and $\gamma_1=-0.016$.
The dark triangular area represents those models which are not of the phantom type for 
$0\le z\le 1$, the top left of the triangle corresponds to $\Lambda$CDM. We see in particular 
that for $\gamma_0<\gamma_0^{\Lambda CDM}$ we get phantom DE for any value of $\gamma_1$.}
\label{fig2}
\end{figure}

Our strategy is therefore simple: Each set ($\gamma_0,\gamma_1$) defines 
$\gamma(z)$ which, through eq.(\ref{eq:all3}), gives in turn the 
background behaviour $\Omega_m(z)$ and therefore $h(z)$. This background can be compared to 
observations which will constrain the set of parameters $(\gamma_0,\gamma_1)$. Usually 
background data, are used to constrain cosmological background parameters like 
$\Omega_{m,0},~\Omega_{\Lambda,0}$ or $\Omega_{DE,0}$, and so on.  
Here, these data are used in order to explore in a systematic way the preferred region 
in the $(\gamma_0,\gamma_1)$ parameter space characterizing the perturbations. 
In this section, we use the Pantheon data \cite{Scolnic:2017caz} consisting of 1048 type Ia supernovae (SNIa) covering the redshift range $0.01 < z < 2.3$, where we have marginalized $\chi^2$ over the parameter $H_0$, results are shown in Fig. \ref{fig2} . We can explore all points in 
the $(\gamma_0, \gamma_1)$ plane around ($\gamma_0^{\Lambda CDM},~\gamma_1^{\Lambda CDM}$), 
for a given value of $\Omega_{m,0}$. We see from Fig. \ref{fig2} that SNIa 
data favour a non constant $\gamma$ which is slightly increasing in time. We also 
see that in a small range of the parameter space $(\gamma_0,\gamma_1)$ non phantom evolution 
can occur in the range $0\le z\le 1$. These points correspond to the triangular area displayed 
on figure \ref{fig2}. 
It is seen in particular that phantomness will always occur in this redshift range for 
$\gamma_0 < \gamma_0^{\Lambda CDM}$. Varying $\Omega_{m,0}$ will affect only marginally the shape 
of the triangle where phantomness is avoided. 

By inspection of the expression for $h(z)$, we see that it is completely fixed once the 
\emph{background} parameter $\Omega_{m,0}$ and the EoS $w(z)$ are given. 
When $h(z)$ is reconstructed from the perturbations, we need the knowledge 
of $\Omega_{m,0}$, and $\gamma(z)$ and hence of $\gamma_0$ and $\gamma_1$ if \eqref{Tx1} 
holds. In that sense, $\gamma(z)$ plays the same role as $w(z)$.    
Note that we can choose freely $\Omega_{m,0}$, $\gamma_0$ and $\gamma_1$ still satisfying 
\begin{align}
\gamma_1 = \frac{1-\Omega_{m,0}}{2\ln\Omega_{m,0}}~\Big[3 w_0 (2\gamma_0-1) +\frac{1+
2\Omega_{m,0}^{\gamma_0}-3\Omega_{m,0}^{1-\gamma_0}}{1-\Omega_{m,0}} \Big]  \lb{gam1}
\end{align}
for $g(0)=1$. While these results are interesting from a mathematical point of view, they imply an 
observational challenge when $\gamma_1$ is much smaller than $\gamma_0$. 
In that case, it will be difficult to measure its value accurately in particular for models 
where it is at the level of $(1-2)\%$. 
In addition, if DE models have their $\gamma_0$ very close to each other,  
$\gamma_0$ would have to be measured with exquisite accuracy in order to differentiate 
these models observationally. 

\begin{figure}
\includegraphics[scale=.7]{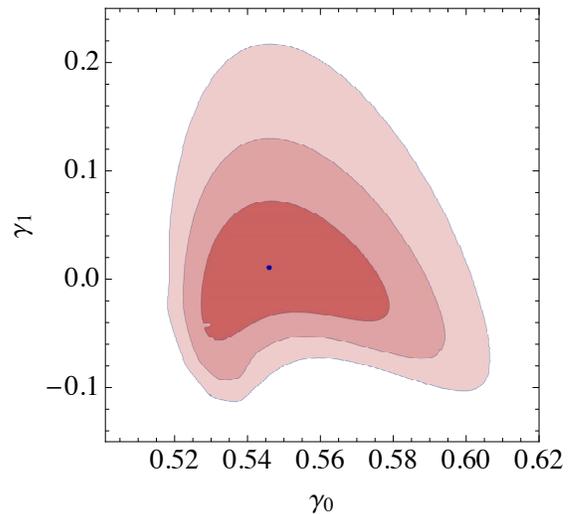}
\caption{The favoured region in the $(\gamma_0, \gamma_1)$ plane is shown when GR is 
assumed $(g=1)$ and $\Omega_{m,0}=0.3$ using cosmic chronometers. These probe directly 
$H(z)$ whence their 
importance. While the presently available data are less constraining, very different 
confidence regions are obtained compared to those resulting from SNIa data.}
\label{fig2b}
\end{figure}

\begin{figure}
\includegraphics[scale=.7]{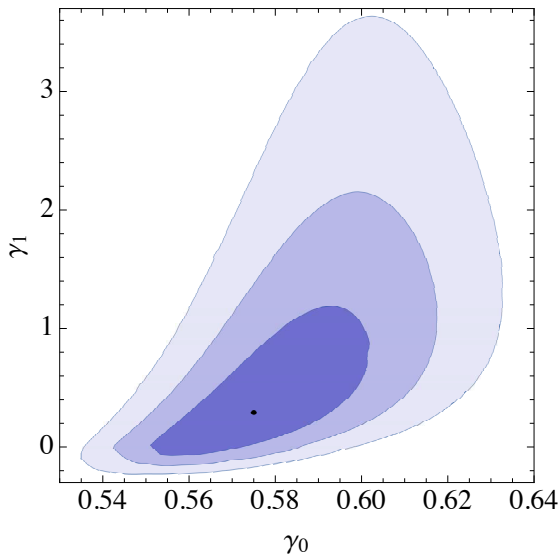}
\caption{The favoured region in the $(\gamma_0, \gamma_1)$ plane is 
shown when GR is assumed $(g=1)$ using $f\sigma_8$ data. It is seen that the constraints 
\emph{today} are substantially weaker than constraints coming from SNIa data 
(see Figure \eqref{fig2}).}
\label{fig2c}
\end{figure}

We also study the parameter space $(\gamma_0,\gamma_1)$ by using 
cosmic chronometers, see Fig.\ref{fig2b}, using data compiled in \cite{Yu:2017iju,Capozziello:2017nbu,Gomez-Valent:2018hwc}. 
While cosmic 
chronometers are presently less accurate than SNIa, they provide a promising way for a 
direct, essentially cosmology independent measurement of $H(z)$ \cite{JL02} 
(see also e.g. \cite{MJVPCC18}) and this is why we find it interesting 
to use them also, however separately. It is interesting that the confidence regions have 
different shapes in parameter space compared with the SNIa confidence regions. 
Finally, we compare the SNIa data to measurements of $f\sigma_8$ 
compiled in \cite{Shafieloo:2018gin}, see Fig.\ref{fig2c}. In that case too, 
SNIa data are much more constraining.
Even if $f\sigma_8$ data provide less constraints on the parameter space $(\gamma_0,\gamma_1)$, 
we can see some sort of tension with SNIa and cosmic chronometers data. It is 
important to note, however, that only 17 data points are considered for $f\sigma_8$ data and 
because they are obtained for a fiducial cosmology, which is different in each survey, these 
data are thus rescaled by the Alckock-Paczinski factor \cite{Alcock:1979mp} 
(see \cite{Nesseris:2017vor}). We stress that the constraints from $f\sigma_8$ data are 
likely to improve substantially in the future.

\begin{figure}
\includegraphics[scale=.7]{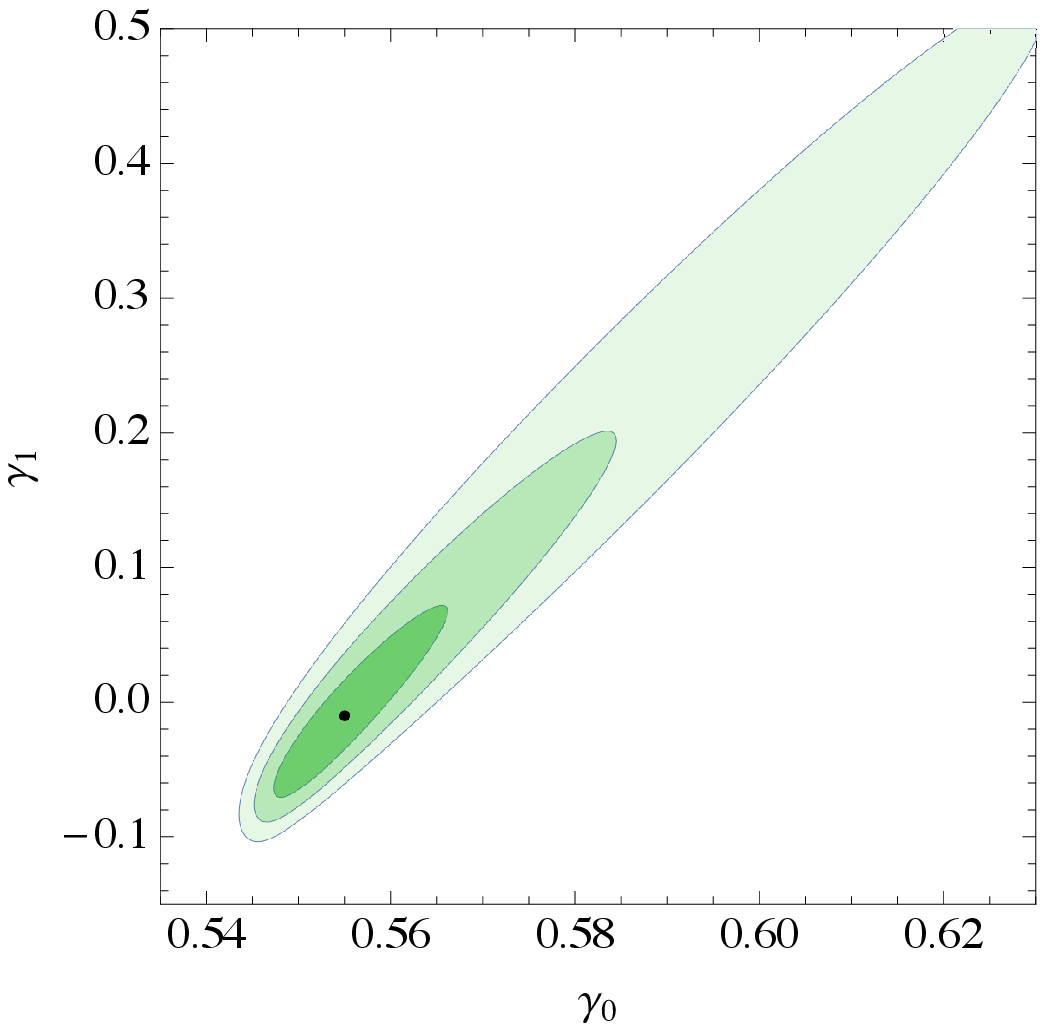}
\includegraphics[scale=.4]{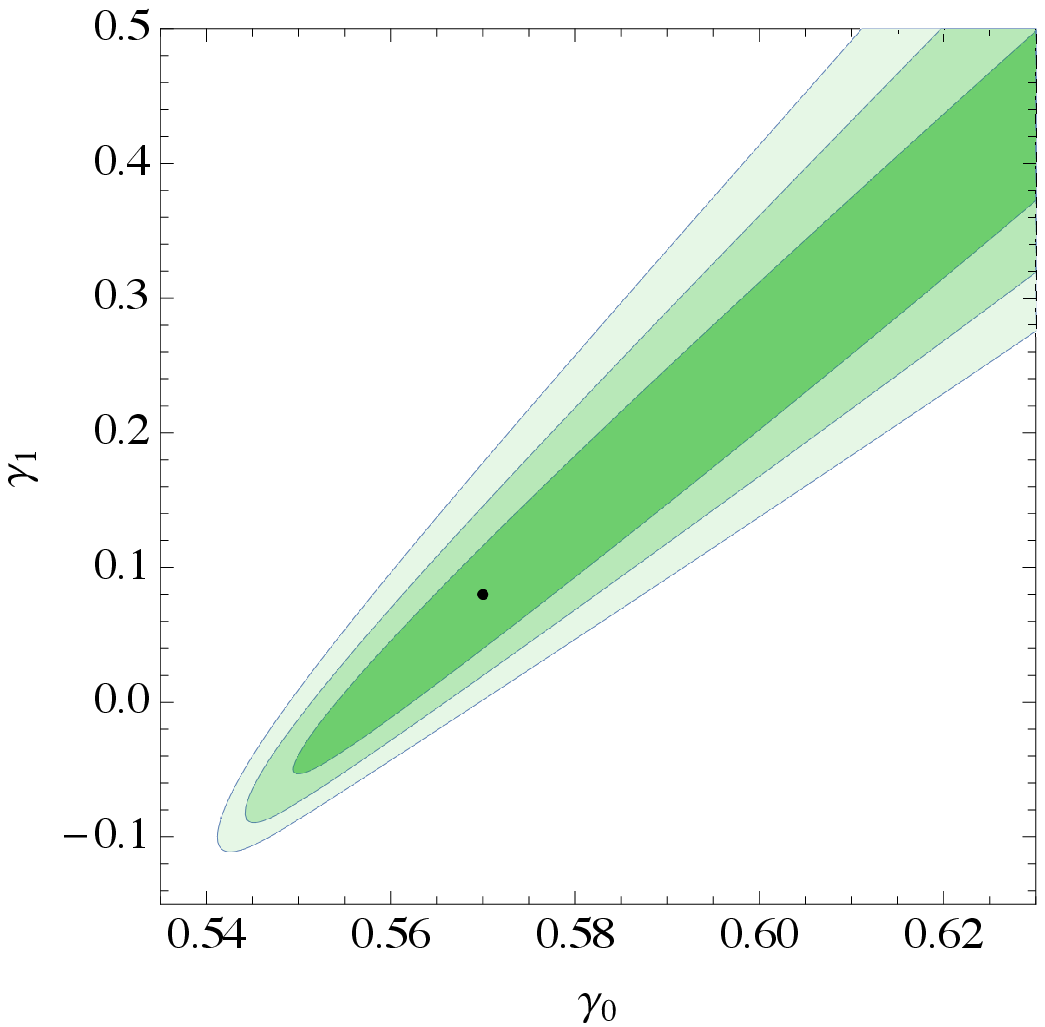}
\includegraphics[scale=.4]{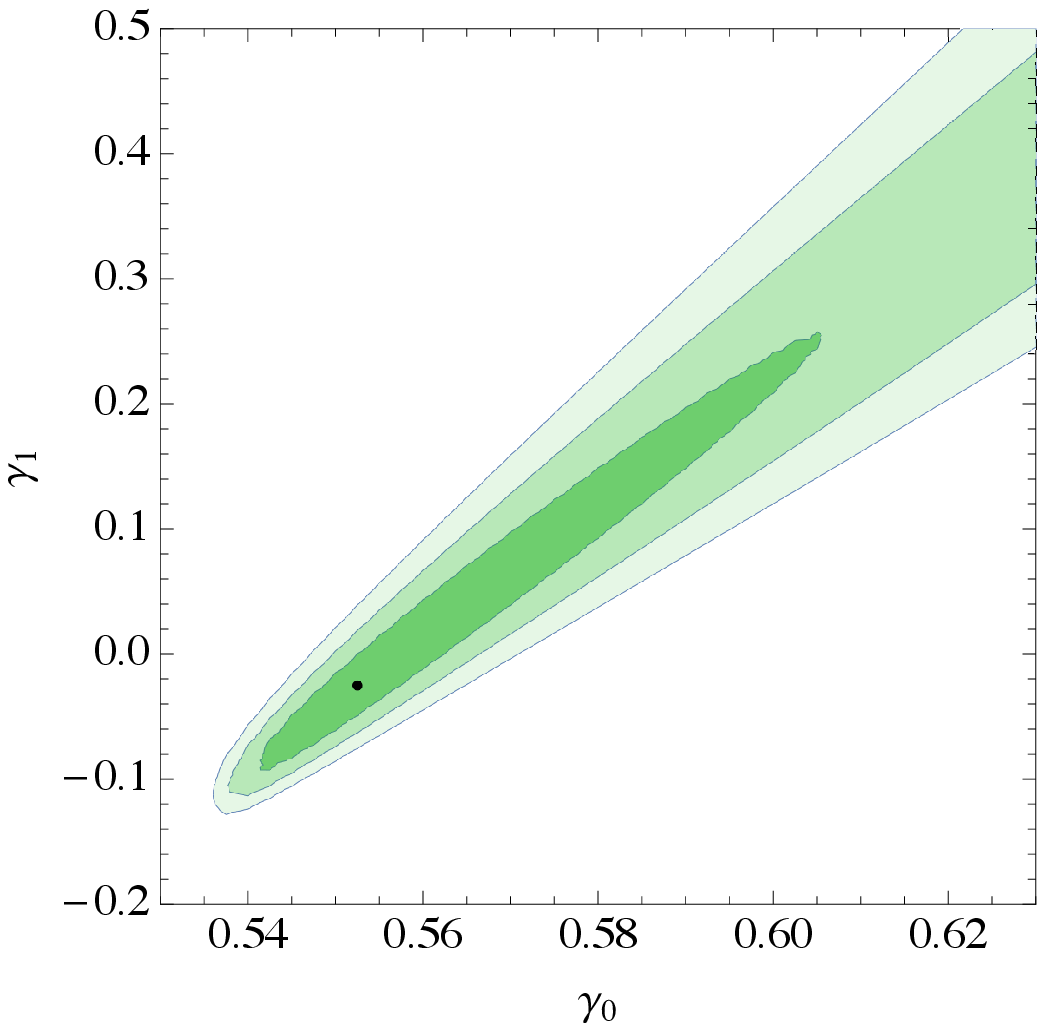}
\caption{The favoured region in the $(\gamma_0, \gamma_1)$ plane is shown when GR is 
assumed $(g=1)$ and $\Omega_{m,0}=0.3$ using data on very low redshifts. 
a) Upper panel: Pantheon SNIa data are used up to $z=0.5$. Lower panel: SNIa data are used up 
to $z=0.35$ for Pantheon data including systematics (on the left) and Union2.1 data \cite{Suzuki:2011hu} without systematics (on the right). On these low redshifts \eqref{Tx1} is a good approximation and the constraints 
derived hold for all models. We see in particular the lower bounds $\gamma_0\gtrsim 0.53$ 
and $\gamma_1\gtrsim -0.15$.}
\label{fig3}
\end{figure}

Considering all constraints, we see that as expected, the constraints on $\gamma_0$ are 
substantially tighter than on $\gamma_1$.
When using SNIa data, models having a quasi-static $\gamma(a)$ with 
$\gamma_1\approx -0.02$ are favoured. While $\Lambda$CDM and noninteracting DE models 
with constant $w_{DE}$ belong to the favoured models, the largest part of the preferred 
region corresponds to phantom DE on low redshifts. 
The preferred region at the $2\sigma$ level lies in the range
$0.549\lesssim \gamma_0\lesssim 0.562$ for $\Omega_{m,0}=0.3$. 
Interestingly, this corresponds 
essentially
to the interval $-1.2\lesssim w_{DE}\lesssim -0.8$ if a constant $\gamma$ is assumed 
\cite{PSG16}. The parameter $\gamma_1$ lies in the range 
$-0.06\lesssim \gamma_1\lesssim 0.05$. 
Hence, while $\gamma_0$ is strongly constrained, larger variation of $\gamma_1$ is allowed.  
Remember that GR is assumed here. 

We recall that these results depend on the assumed behaviour \eqref{Tx1}. We can try to derive 
results which are essentially model-independent by using data only on very small redshifts so 
that \eqref{Tx1} now serves as a good fit. Though constraints necessarily become less 
stringent, conclusions drawn on the other hand are more general. We see from the lower panel 
of figure \ref{fig3} that, at the $3\sigma$ confidence level, models with 
$\gamma_1\lesssim -0.15$ or with $\gamma_0\lesssim 0.53$ cannot be obtained inside GR.    
These results are in agreement with results obtained earlier for $f(R)$ models. 

The constraints on $\gamma_0$ and $\gamma_1$ were obtained using the background expansion. 
As these data are expected to remain more accurate than perturbations data, so are the 
inferred constraints on $\gamma_0$ and $\gamma_1$. We insist here again that the 
reconstructed function $h(z)$ is a genuine theoretical prediction.  
Another interesting aspect is connected to the value of $H_0$. Indeed, $\gamma_0$ and 
$\gamma_1$ yield a reconstruction of $h(z)$, not of $H(z)$. Hence a pair $\gamma_0$ and 
$\gamma_1$, and therefore the underlying model, can be in tension with $H(z)$ data, and 
even ruled out, depending on the $H_0$ value which is assumed. In our analysis we have chosen 
to marginalize the data over $H_0$. 
The results of this subsection do not exclude the well-known possibility to distinguish models 
with significantly different $\gamma_0$ (and necessarily larger $\gamma_1$). 
Among the appealing cases where this can happen are modified gravity DE models to which we 
turn our attention now.  

\subsection{Reconstruction of g}
We want to explore now another useful reconstruction. It was soon realized that 
a host of models are able to produce an accelerated expansion and even to produce an 
expansion rate close to that of $\Lambda$CDM. Hence, it is reasonable to assume some $h(z)$, 
which can later be refined as more accurate data will be released and to explore the 
possible behaviours of $\gamma(z)$, giving the matter perturbations, and of $g(z)$ which 
encodes the gravitational force driving these perturbations. We can use the 
expansion \eqref{Tx} up to first order around the present time which yields a good 
approximation on very low redshifts up to $z\lesssim (0.35-0.5)$.  
So we will assume a background evolving like $\Lambda$CDM and take $\gamma$ given by 
\eqref{Tx1}. In this framework, we can reconstruct the evolution of 
$g(z)$, and this reconstruction will be accurate on all redshifts where \eqref{Tx1} holds. 

\begin{figure}
\includegraphics[scale=.7]{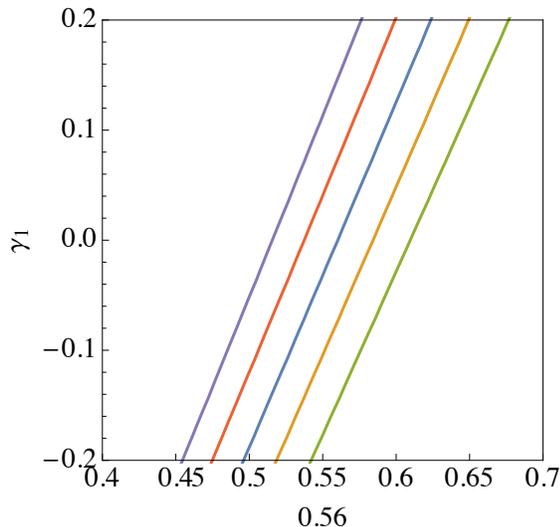}
\caption{The background is fixed to our fiducial $\Lambda$CDM 
model with $\Omega_{m,0}=0.3$.
From the left to the right, lines correspond to 
$g(0)=1.2,~1.1,~1,~0.9,~0.8$. Points on the left, resp. on the right, of each line yield 
a higher, resp. lower, $g(0)$. Hence the line $g(0)=1$ divides 
the plane in models with $g(0)>1$ (upper part) and $g(0)<1$ (lower part). 
Equivalently, if we fix the value of $\gamma_0$, increasing, resp. decreasing, $\gamma_1$ 
will increase, resp. decrease, $g(0)$. 
}
\label{fig4a}
\end{figure}

Once the background evolution is known and some ansatz is used for $\gamma(z)$, 
$g(z)$ is solved algebraically from \eqref{eq:all3}. A first important point concerns the 
present-day value $g(0)$. Inspection of \eqref{eq:all3} shows that 
$\gamma_1$ contribute to its determination, raising its value for positive 
$\gamma_1$ and lowering it for negative $\gamma_1$. On figure \ref{fig4a}, points 
$(\gamma_0, \gamma_1)$ corresponding to constant $g(0)$ are shown and it is seen that 
they correspond to straight lines. 
Any of these lines divide the plane in such a way that the domain on the left corresponds 
to a higher $g(0)$ while the domain on the right corresponds to a lower $g(0)$. Of particular 
importance is the line corresponding to $g(0)=1$, for which the effective gravitational 
constant equals (the GR) Newton's constant $G$ today. Modified gravity models that cannot allow 
for $g<1$ are excluded from the domain on the right of this line. 

\begin{figure}
\includegraphics[scale=.7]{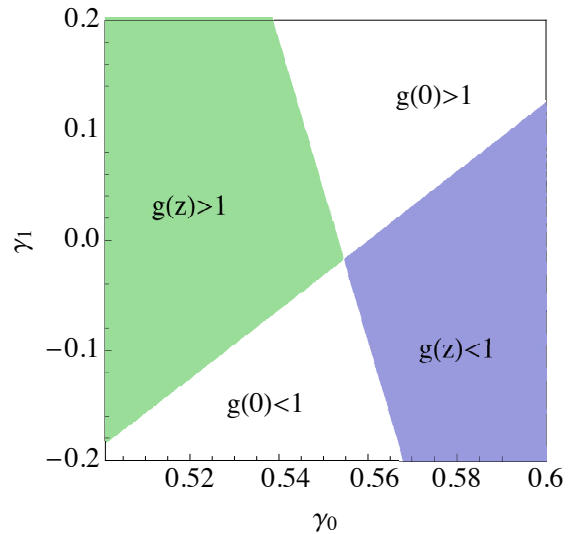}
\caption{The background is fixed to our fiducial $\Lambda$CDM 
model with $\Omega_{m,0}=0.3$.
Each pair $\gamma_0,~\gamma_1$ 
represents a modified gravity DE model with fixed $\Lambda$CDM background evolution. The 
plane around $(\gamma_0^{\Lambda CDM},~\gamma^{(1),\Lambda CDM})$ is divided into four regions. 
To the left of $(\gamma_0^{\Lambda CDM},~\gamma^{(1),\Lambda CDM})$ are those models with $g(z)>1$, 
to the right those with $g(z)<1$. Models inside the upper (inverted) triangle start with 
$g(0)>1$ and later cross the value $g=1$ downwards for $z < 0.35$, models inside the lower 
triangle start with $g(0)<1$ and cross $g=1$ upwards for $z < 0.35$.}
\label{fig4b}
\end{figure}

However the subsequent behaviour for $z>0$ can lead to a crossing of this line. Studying the 
behaviour of $g(z)$ in the $(\gamma_0, \gamma_1)$ plane, still assuming a fixed 
$\Lambda$CDM background, we find the structure shown on figure \ref{fig4b}. 

The area around the point corresponding to $\Lambda$CDM can be divided into four regions. 
An inverted triangle is found above ($\gamma_0^{\Lambda CDM},~\gamma_1^{\Lambda CDM}$) where 
$g$ started above one, $g(0)>1$, and later satisfies $g(z)<1$ at the redshift $z=0.35$. Hence 
for points inside this triangle, the effective gravitational constant has crossed downwards 
the value $G$ in the interval $0 < z < 0.35$.   
A similar triangle is found below ($\gamma_0^{\Lambda CDM},~\gamma_1^{\Lambda CDM}$) with the 
opposite behaviour, $g(0)<1$ and $g(z)>1$ at $z=0.35$. In the remaining region on the left of 
$(\gamma_0^{\Lambda CDM},~\gamma_1^{\Lambda CDM})$ with $\gamma_0 < \gamma_0^{\Lambda CDM}$, one has 
$g>1$ always up to $z=0.35$, while in the remaining region on the right of 
$(\gamma_0^{\Lambda CDM},~\gamma_1^{\Lambda CDM})$ with $\gamma_0 > \gamma_0^{\Lambda CDM}$ we obtain 
$g<1$ always up to $z=0.35$. The left side of the upper triangle and the right side of the 
lower triangle represent those points for which $g=1$ at $z=0.35$. The line with the opposite 
sides of the triangles are those points starting with $g(0)=1$.

We note here an interesting mathematical property. 
Assuming that our ansatz for $\gamma(z)$ holds for large $z$ too, 
though we emphasize that this is generically not the case for modified gravity models, 
we can extend the figure for $z\to \infty$. As we move to higher redshifts, more and more 
models will cross the value $g=1$, either downwards in the upper triangle, or upwards in the 
lower triangle. The left side of the upper triangle will move 
(counterclockwise) slightly to the left, and the right side of the lower triangle will move 
slightly to the right. The limit will be given by the line 
$\gamma_0 + \gamma_1=\frac{6}{11}$.
\begin{figure}
\includegraphics[scale=.7]{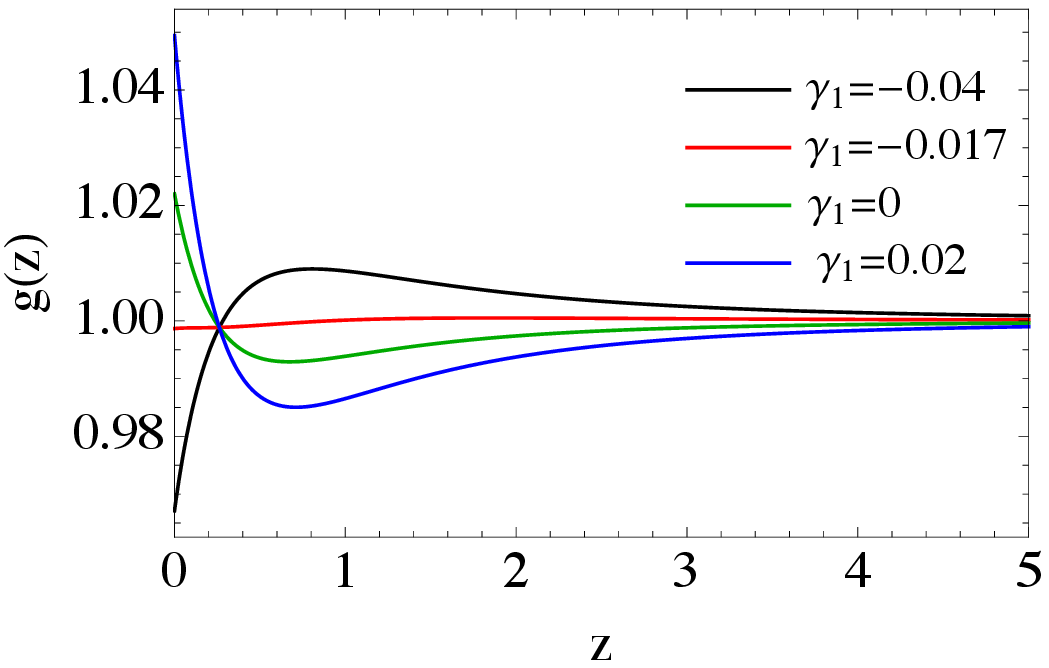}
\includegraphics[scale=.7]{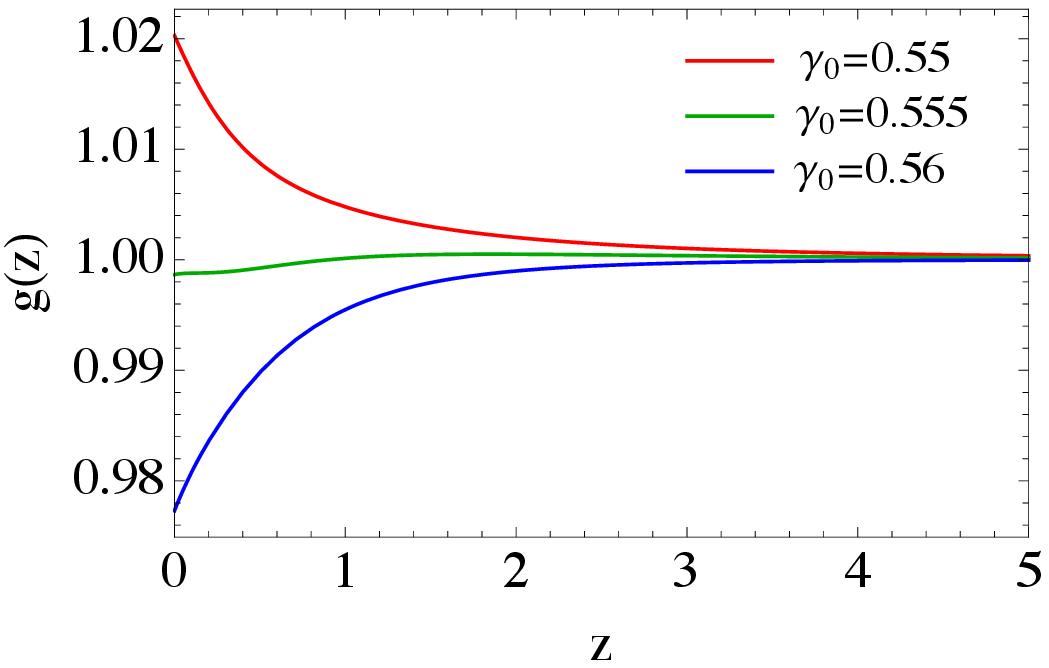}
\caption{a) The figure on the upper panel shows the behaviour of $g(z)$ when we move in the 
$\gamma_0-\gamma_1$ plane above and below the fiducial $\Lambda$CDM model, 
$\gamma_0=0.555,~\gamma_1=-0.017$ and $\Omega_{m,0}=0.3$. When we move 
upwards $(\gamma_1>\gamma_1^{\Lambda CDM})$, resp. downwards 
($\gamma_1<\gamma_1^{\Lambda CDM}$), $g(0)$ increases, resp. decreases, and $g(z)$ crosses 
the value $g=1$ at small redshifts.   
b) On the lower panel, the behaviour of $g(z)$ is shown when we move in the 
$(\gamma_0,\gamma_1)$ plane to the left and to the right of our fiducial 
$\Lambda$CDM model. A moderate departure from $g=1$ is obtained with $g>1$, resp. $g<1$, on 
the left, resp. right, of $\gamma_0^{\Lambda CDM}$.}  
\label{fig6}
\end{figure}
Indeed, the asymptotic behaviour of $g(z)$ for large $z$ at arbirary points in the 
$(\gamma_0, \gamma_1)$ plane is given by 
\be
g(z) \sim 1 + \frac{1-\Omega_{m,0}}{3 \Omega_{m,0}~z^3}~
                                   \left[6 - 11(\gamma_0 + \gamma_1) \right]~. \lb{ginf}
\ee
In the upper triangle we have $g\to 1$ from above, hence the last term in \eqref{ginf} tends 
to zero while positive; in the lower triangle we have the opposite situation and the last 
term of \eqref{ginf} tends to zero while negative. This is possible for 
$\gamma_0 + \gamma_1=\frac{6}{11}$ only. Another way to see this is as follows. 
We know from theoretical considerations that the limit 
$\gamma_{-\infty} = \frac{6}{11}$ is obtained for $\Lambda$CDM in the asymptotic past 
\cite{PSG16}. However the line corresponding to $g=1$ at $z\to \infty$ corresponds just to the 
$\Lambda$CDM model itself at $z\to \infty$, hence it must 
satisfy $\gamma_{-\infty} = \gamma_0 + \gamma_1 = \frac{6}{11}$. 

Models with $\gamma_0\approx \gamma_0^{\Lambda CDM}$ below the $1\%$ level cannot be clearly 
differentiated from $\Lambda$CDM with measurements of $\gamma_0$ as long as $\gamma_1$ 
is not measured accurately, which can be expected for $\Big|\gamma_1\Big|\lesssim 0.05$. 
If we move along $\gamma_0$ to the right or to the left of 
$(\gamma_0^{\Lambda CDM},~\gamma_1^{\Lambda CDM})$, 
we get models with $g(0)$ departing moderately from $1$ on very small redshifts, see 
right panel of figure \ref{fig6}. 
If we move upwards, resp. downwards, along $\gamma_1$ inside the upper, resp. lower, 
triangle, we have models mimicking $\Lambda$CDM for $\gamma_1$ not too large with $g(0)$ 
substantially higher, resp. lower than one. 
Those models however necessarily cross the value $g=1$ at very low redshifts, 
hence they cannot be realized in models not allowing for such a crossing. 

To summarize, models that cannot be distinguished observationnally from $\Lambda$CDM through 
the measurement of $\gamma_0$ alone are mostly modified gravity models which can depart 
substantially from GR but which must necessarily allow for a crossing of $g=1$ on small 
redshifts.\\

\section{Conclusions}
A large family of noninteracting DE models inside GR, with $\Lambda$CDM among them, exhibits a  
quasi-constant behaviour of the growth index $\gamma(z)$.
For these models, the  behaviour \eqref{Tx1} of $\gamma(z)$ can be 
expressed with two parameters only, namely $\gamma_0$ and $\gamma_1$.
It is then possible to reconstruct $h(z)$ for these models using 
the parameters $\gamma_0$ and $\gamma_1$. Motivated by these examples, we have constrained 
systematically the two parameters $\gamma_0$ and $\gamma_1$ using background expansion data
for all models satisfying \eqref{Tx1} and we have found the preferred region in the 
$(\gamma_0, \gamma_1)$ plane. We have obtained that while $\gamma_0$ is rather tightly 
constrained around $\gamma_0^{\Lambda CDM}$, a large range remains for the parameter 
$\gamma_1$. Such an accuracy could not be obtained using perturbations data in view of 
the large errors on the growth function $f$. 
We have refined our analysis by using background data on very small redshifts, so that the 
assumed behaviour \eqref{Tx1} becomes a good approximation for all (reasonable) models. 
We find in particular that $\gamma_0$ and $\gamma_1$ are bounded from below. 
Values measured below these bounds, and such models were found earlier, would hint at either 
modified gravity (see e.g. \cite{MG},\cite{KP18}) or interacting DE models (see e.g. \cite{IDE,CalderaCabral:2009ja}). 

We have also considered DE modified gravity models assuming a fixed fiducial background 
dynamics, $\Lambda$CDM in our analysis. Though a quasi-constant 
behaviour for $\gamma(z)$ cannot be assumed in this case, it can be used on very small 
redshifts, in the important range where DE is expected to induce the universe present 
accelerated expansion. 
We have investigated modified gravity models which cannot be discriminated from $\Lambda$CDM 
as a result of large errors on the parameter $\gamma_1$ while observations are able to 
pinpoint the value of $\gamma_0$ below the $1\%$ level. Though some investigated modified 
gravity models yield a significantly lower $\gamma_0$, we study here the price to pay for 
modified gravity models in order to satisfy $\gamma_0\approx \gamma_0^{\Lambda CDM}$.
We have found that models with a substantial variation of the effective 
gravitational coupling today will cross Newtons constant $G$ on very small redshifts, either 
upwards or downwards. This gives a very strong constraint, further restraining modified gravity models able to realize this phenomenology. For example, 
$f(R)$ DE models do not allow for such a behaviour though this is possible in other 
models \cite{FHKMTZ16}. As the goal of future experiments will be to probe 
the growth index and a possible departure from GR, a systematic study of the consistency of 
the background expansion with the perturbations along the lines presented in this work can 
give interesting phenomenological constraints as well as new insights. 
 

\section*{Acknowledgments}

The authors thank Benjamin L'Huillier for many helpful comments and for sharing with us compiled $f\sigma_8$ data used in this paper. The work of R. Gannouji is supported by Fondecyt project No 1171384.


\end{document}